# Quantum Hall Effects in Monolayer-Bilayer Graphene Planar Junctions

*Jifa Tian[1,2,*], Yongjin Jiang[3,1], Isaac Childres[1,2], Helin Cao[1,2], Jiangping Hu[1], and Yong P. Chen[1,2,4,*]*

1. Department of Physics, Purdue University, West Lafayette, Indiana 47907, USA
2. Birck Nanotechnology Center, Purdue University, West Lafayette, Indiana 47907, USA
3. Center for Statistical and Theoretical Condensed Matter Physics, and Department of Physics, Zhejiang Normal University, Jinhua 321004, P. R. China
4. School of Electrical and Computer Engineering, Purdue University, West Lafayette, Indiana 47907, USA

*Emails: tian5@purdue.edu; yongchen@purdue.edu

## Abstract

The Hall resistance of a homogeneous electron system is well known to be anti-symmetric with respect to the magnetic field and the sign of charge carriers. We have observed that such symmetries no longer hold in planar hybrid structures consisting of partly single layer graphene (SLG) and partly bilayer graphene (BLG) in the quantum Hall (QH) regime. In particular, the Hall resistance ($R_{12xy}$) across the SLG and BLG interface is observed to exhibit quantized plateaus that *switch* between those characteristic of SLG QH states and BLG QH states when either the sign of the charge carriers (controlled by a back gate) or the direction of the magnetic field is reversed. Simultaneously reversing both the carrier type and the magnetic field gives rise to the same quantized Hall resistances. The observed SLG-BLG interface QH states, with characteristic asymmetries with respect to the signs of carriers and magnetic field, are determined only by the *chirality* of the QH edge states and can be explained by a Landauer-Büttiker analysis applied to such graphene hybrid structures involving two regions of different Landau level (LL) structures.

The success of isolating single layer carbon lattices (graphene) has opened an exciting research field for both fundamental physics and potential nanoelectronic devices applications.[1-5] Graphene exhibits an unusual band structure showing a linear relationship between the energy and momentum near the Dirac point and allowing electric field tuning of both the type and the density of charge carriers, which are massless Dirac fermions capable to reach impressive high mobilities [1,6-9]. Previous studies have demonstrated that SLG shows half-integer quantum Hall effects (QHE), where the Hall resistance is quantized at values of $R_{xy}=h/v_1e^2$ around filling factors $v_1 = \pm 4(N+1/2)$.[2,3] Here, $N$ is an integer, $e$ is the electron charge, $h$ is Planck's constant and the factor of four is due to spin and valley degeneracy. Bilayer graphene (BLG), composed of two graphene monolayers weakly coupled by interlayer hopping, is also interesting and has been shown to have additional unique physical properties than those of SLG.[10-12] Charge carriers in BLG (assuming normal AB stacking) are massive *chiral* fermions with a Berry's phase $2\pi$.[10] BLG is observed to display integer QHE of quantized Hall plateaus $R_{xy}=h/v_2e^2$ around filling factors $v_2=\pm 4N$.

It is well know that the Hall resistance ($R_{xy}$) of a homogenous electron system is anti-symmetric with respect to carrier density ($n$, where positive $n$ refers to holes, negative $n$ refers to electrons) and magnetic field ($B$), i.e., $R_{xy}(-n/B)=-R_{xy}(n/B)$. Such electron-hole (*e-h*) and $B$ symmetries in the Hall transport are also obeyed in QHE observed in 2D electron systems (2DES) in general [13], including the QHE for both SLG and BLG. [2,3,10] We will show in this paper that such a familiar symmetry is violated in a remarkable way in a SLG-BLG hybrid planar structure (even when the carrier density $n$ may be uniform).

A SLG-BLG hybrid planar structure consists of partly SLG and partly BLG. As the SLG and BLG parts possess qualitatively different electronic band structures, such a planar heterostructure between two segments with completely different LL sequences (and different Berry's phases for charge carriers [5]) is fundamentally interesting with no analog in conventional 2DES and may result in novel physics and transport properties. While the QHE of either SLG or BLG has been extensively studied, hybrid SLG-BLG structures [14-18] have received relatively little attention. A recent theoretical work has predicted rich structures in the interface LL in a SLG-BLG junction, but its transport properties have not been studied.[15] Experimentally, C. P. Puls *et al.* studied the quantum oscillations and observed various anomalous features in two-terminal magnetoresistance *parallel* to the interface of hybrid SLG-BLG structures.[14] Unusual features in the two-terminal magnetoconductance [17] and a longitudinal resistance ($R_{xx}$) asymmetric in the signs of carriers and magnetic field have been observed cross SLG-BLG interfaces [18]. To our knowledge, however, there is no experimental work reported on the interface QHE (especially the Hall resistance) in a SLG-BLG hybrid structure. Here, we report a systematic study of QH transport in such junctions and the observation of *interface* QH states, where none of the usual *e-h* and *B* symmetries are obeyed. The interface QH plateaus are found to switch between SLG and BLG like values when either *n* or *B* is reversed. We also present a theoretical model using a Landauer-Büttiker analysis to explain the main observations.

Our hybrid structures, consisting of partial SLG and BLG, are mechanically exfoliated from HOPG (ZYA, Momentive Performance Materials Quartz Inc) and transferred onto a wafer with 280 nm thermal oxide ($SiO_2$) on top of a p++ Si substrate [19], and selected by the optical contrast [20] and Raman spectra [21] of SLG and BLG flakes. A typical optical image of a hybrid SLG-BLG structure is shown in Fig. 1a. Some representative Raman spectra of the SLG and BLG parts are shown in Fig. 1b. We use a standard e-beam lithography process [19] to fabricate devices with various geometries. Two devices ("1" and "2", shown in Figs. 2a and 3a), will be presented in this work. The measurements are performed in a helium-3 system with B up to ± 18 T. At B=0T and T=0.5K, we performed field effect measurements (4-terminal resistance, R, versus back gate voltage, $V_g$) on both the SLG and BLG parts in both devices "1" and "2". The corresponding results are shown in Fig. 1c,d, respectively. In device "1", the CNPs are $V_{1CNP} \sim 10V$ and $V_{2CNP} \sim 17$ V for the SLG and BLG parts, respectively. For device "2", the measured $V_{CNP}$ of both the SLG and BLG parts are ~20V. Note there can be slight hysteresis and variations (~2V) in $V_{CNP}$ in our devices for different gate sweeps and measurements.

Figures 2a,b show the optical image and corresponding schematic 3D structure of device "1", where SLG and BLG are connected in series between the drain ("D") and source ("S") electrodes. In this device, the Hall and longitudinal resistances of the SLG part ($R_{1xy}$ and $R_{1xx}$), BLG part ($R_{2xy}$ and $R_{2xx}$), and SLG-BLG interface ($R_{12xy}$ and $R_{12xx}$) can all be measured simultaneously. The measured low field magnetoresistance and Hall resistances of both SLG and BLG parts are also used to extract their carrier densities and Hall mobilities. At $V_g$=0V, the carrier (hole) density of the SLG part ($p_{SLG}$) is ~9×10$^{11}$ cm$^{-2}$ and the Hall mobility ($\mu_{SLG}$) is ~3100 cm$^2$/Vs. The carrier density of BLG ($p_{BLG}$) part is ~1.8×10$^{12}$ cm$^{-2}$ and the mobility ($\mu_{BLG}$) is ~1800 cm$^2$/Vs. Figure 2c shows the Hall resistances ($R_{1xy}$, $R_{2xy}$ and $R_{12xy}$) and longitudinal resistances ($R_{1xx}$, $R_{2xx}$ and $R_{12xx}$) as functions of back gate voltages ($V_g$) at $B$ = +15 T. The corresponding edge state chiralities [13], either clockwise (CW) or anticlockwise (ACW), are labeled near representative QH plateaus. In SLG part, we observe a series of well-developed QH plateaus in $R_{1xy}$ at $\pm h/2e^2$, $\pm h/6e^2$, $\pm h/10e^2$, where the corresponding $R_{1xx}$ are vanishing. We also observe QH states from the BLG part with $R_{2xy}$ quantized at $\pm h/4e^2$, $h/8e^2$, and $h/12e^2$. These results indicate that the characteristic QHE observed in SLG or BLG are still preserved in the SLG and BLG parts of the graphene hybrid structures, and obeys *e-h* symmetry. Interestingly, the Hall resistance $R_{12xy}$ of the SLG-BLG interface is also observed to show quantized plateaus. In particular, when the charge carriers are holes ($V_g$<~10V), $R_{12xy}$ is observed to closely resembles BLG $R_{2xy}$, and displays a well-developed plateau at $h/4e^2$, as well as other developing plateaus near $h/8e^2$ and $h/12e^2$, all corresponding to the values of BLG QH states. However, when the carriers are changed to electrons for $V_g$>~20V, $R_{12xy}$ shows a developing plateau close to $-h/2e^2$, a value corresponding to a SLG QH state. Alternatively, we can also measure the QHE by tuning $B$ at a fixed $V_g$. Figure 2d shows the results of such measurements $V_g$= 0 V (where the carriers are holes in both parts). The Hall resistance $R_{1xy}$ ($R_{2xy}$) of the SLG (BLG) part exhibits characteristic SLG (BLG) QH states near ±15T with well-defined plateaus at $\pm h/2e^2$

($\pm h/4e^2$) accompanied by vanishing $R_{1xx}$ ($R_{2xx}$), showing that the QHE measured in the SLG and BLG parts are individually symmetric with $B$. The interface $R_{12xy}$ also shows well defined quantized Hall plateaus, however they are *asymmetric* upon the reversal of the $B$. Near +15 T, $R_{12xy}$ shows a quantized plateau at *h/4e²* corresponding to a BLG QHE, accompanied by a vanishing $R_{12xx}$. On the other hand, near -15 T, $R_{12xy}$ is quantized at *-h/2e²* (value for a SLG QHE) with accompanying $R_{12xx}$ now displaying a resistance plateau (at *h/4e²*) rather than vanishing. The observations above (Figs 2c and 2d) that the quantized $R_{12xy}$ switch between SLG-like and BLG-like when *either* the carrier type or the $B$ field is reversed implies that simultaneous reversing *both* the carrier type *and* $B$ field would give rise to the same quantized $R_{12xy}$ (e.g. the SLG-like QH plateau of *-h/2e²* seen for both electrons, +15T in Fig. 2c and holes, -15T in Fig. 2d). Our results also suggest that the interface QH states and their quantized Hall plateau values are determined by the *chirality* of edge state currents (e.g., in Figs. 2c,d, the BLG-like QH plateaus are observed with ACW edge state chirality whereas the SLG-like QH plateau (*-h/2e²*) is observed with CW edge state chirality).

Qualitatively similar results are also observed in device "2" (Figs. 3a,b). Here the "D" electrode connects to both the SLG and BLG parts and the "S" electrode is touching the SLG only, and we mainly focus on the Hall resistance $R_{12xy}$ crossing the interface (measured between electrodes "c" and "b"). The measured $V_{CNP}$ of both the SLG and BLG parts are ~20V. Due to its structure and electrodes configuration, device "2" does not allow measurement of individual Hall effects from its SLG and BLG parts. We can extract an effective (average) carrier (hole) density ~$1.8\times10^{12}$ cm$^{-2}$ from low-B $R_{12xy}$ and a mobility (µ, using $R_{xx}$ measured between "a" and "b" in the SLG part) to be ~7200 cm²/Vs at $V_g$=0. In Fig. 3c, we again see that $R_{12xy}$ (as function of $V_g$ at B = ±18 T) shows quantized plateaus that switch between SLG-like and BLG-like QH plateau values when reversing the sign of carriers. For example, when *B = -18* T, $R_{12xy}$ exhibits plateaus at *-h/10e²*, *-h/6e²*, *-h/2e²* (values of SLG QH states) for $V_g$< $V_{CNP}$ (holes), but at *+h/4e²*, *+h/8e²* (values of BLG QH states) for $V_g$>$V_{CNP}$ (electrons). By reversing (thus changing the chirality of the edge state currents) the $B$ field, the relatively well developed quantized Hall plateaus of $R_{12xy}$ now switch from BLG-like QH plateaus (+$h/12e^2$, +$h/8e^2$, +$h/4e^2$) to SLG-like QH plateaus (-$h/2e^2$ and -$h/6e^2$) as carriers switch from holes to electrons. It is also notable that the values of the relative well developed SLG-like QH plateaus in $R_{12xy}$ tend to be negative (CW edge state chirality), whereas those corresponding to BLG-like QH states tend to be positive (ACW edge state chirality). This is again confirmed by the $B$-dependent $R_{12xy}$ shown in Fig. 3d. In particular, we observe the SLG-like plateau developing at -$h/2e^2$ for holes ($V_g$ = 0 V < $V_{CNP}$) near B=-18 T and well-developed for electrons ($V_g$ = 30 V > $V_{CNP}$) near B=+18 T (both CW edge state chirality), and the BLG-like plateau developing at +$h/4e^2$ for holes ($V_g$ = 0 V) nears +18 T and well-developed for electrons ($V_g$=30 V) near -18 T (both ACW edge state chirality). We also note in Fig. 3d that at lower B, $R_{12xy}$ can show developing plateaus of both SLG-like and BLG-like values (as labeled in the figure) even with the same carrier type and B direction (e.g. for $V_g$=30 V, electrons, and B<0, we see plateaus around $h/10e^2$ and $h/14e^2$, corresponding to SLG-like QH plateau values, in addition to BLG-like $h/4e^2$, $h/8e^2$). Such a "mixed" appearance is not fully understood and possible reasons will been discussed later.

The main features of the observed SLG-BLG interface QHE can be understood using a Landauer-Büttiker [22,23] analysis of the chiral edge states. We first consider a generic hybrid structure, consisting of two regions at different QH states (Fig. 4a,b). A similar model has been analyzed for a 4-terminal SLG p-n junction in Ref. 24. Figs. 4a,b show measurement schematics with CW and ACW edge currents, respectively. Here we assume that the two regions are each in a QH state with the number of edge states $m_1$ and $m_2$ (both regions having quantized Hall resistances $R_{SD,ae}=(V_a-V_e)/I_{SD}=h/m_1e^2$ and $R_{SD,cb}=h/m_2e^2$, with $m$ < 0 for CW edge state chirality and $m$ > 0 for ACW chirality). The source-drain current $I_{SD}$ is:

$$I_{SD} = \frac{e^2}{h}(V_S - V_D) \cdot \min(|m_1|,|m_2|) = \frac{e^2}{h}(V_S - V_D) \cdot |m_1| \qquad (1)$$

(we here assume $m_1 \cdot m_2 > 0$ and $|m_2| > |m_1|$, corresponding to the most typical situation in our SLG-BLG hybrid structures, see also Fig. 4c) [25]. The only difference between Fig. 4a and Fig. 4b is the edge state chirality, determined by the sign of charge carriers and direction of the *B* field (reversing either of them reverses the chirality, while reversing both leads to the same chirality). In this case, either electrons under +*B* or holes under –*B* give CW edge currents, whereas electrons under –*B* or holes under +*B* give ACW edge currents (note that the QH edge state chirality is opposite to the chirality of the cyclotron orbits [22,23]). Using a Landauer-Büttiker analysis, one can calculate [24] the longitudinal resistance for the interface to be:

$$R_{SD,ca} = \frac{V_c - V_a}{I_{SD}} = \begin{cases} (1/m_2 - 1/m_1)h/e^2, & CW \\ 0, & ACW \end{cases} \quad (2a)$$

$$R_{SD,be} = \frac{V_b - V_e}{I_{SD}} = \begin{cases} 0, & CW \\ (1/m_1 - 1/m_2)h/e^2, & ACW \end{cases} \quad (2b)$$

On the other hand, the interface Hall resistances are,

$$R_{SD,ab} = \frac{V_a - V_b}{I_{SD}} = h/(\max(m_1, m_2)e^2) = \begin{cases} h/(m_1 e^2) = -h/|m_1|e^2, & CW \\ h/(m_2 e^2) = h/|m_2|e^2, & ACW \end{cases} \quad (3a)$$

$$R_{SD,ce} = \frac{V_c - V_e}{I_{SD}} = h/(\min(m_1, m_2)e^2) = \begin{cases} h/m_2 e^2 = -h/|m_2|e^2, & CW \\ h/m_1 e^2 = h/|m_1|e^2, & ACW \end{cases} \quad (3b)$$

Note Eqs. (2b&3b) can also be obtained from Eqs. (2a&3a) by reflecting the device with respect to the "S"-"D" axis [25]. We also note that the interface Hall resistance $R_{12xy}$ can also be obtained from interface longitudinal resistance (Eq. (2)) as $R_{12xx}$ + $R_{xy}$(SLG/BLG) (for example $R_{SD,ab}=R_{SD,ac}+R_{SD,cb}$).

Now we apply these analyses to the SLG-BLG hybrid structure, with the SLG and BLG parts corresponding to regions "1" and "2", respectively. The configuration of device "1" (Fig. 2) is the same as that shown in Figs. 4a,b. Eq. (2a) agrees well with the observed interface $R_{12xx} = h/4e^2$ at B=-15T (where $m_2$ = -4, $m_1$ = -2, with CW circulation) as well as $R_{12xx} = 0$ at B=15T ($m_2$ = 4, $m_1$ =2, ACW circulation) in Fig. 2d.

Eqs. (3a,b) for a SLG-BLG hybrid structure can be summarized in Fig. 4c showing the schematic illustration of the QHE in SLG, BLG, and SLG-BLG interface (assuming a SLG-BLG hybrid structure with uniform $n$ or $v$). It can be seen that $|m_2|$ is always larger than $|m_1|$ at the same filling factor. It also clearly shows the edge chirality induced switching between SLG-like and BLG-like QH plateau values in the interface quantized Hall resistances. For $R_{SD,ab}$ (corresponding to experimentally measured $R_{ch}$ in device "1"), the CW edge chirality gives rise to SLG-like QH plateaus whereas ACW chirality gives rise to BLG-like QH plateaus. All these are again in agreement with the experimental observations (see Fig. 2c,d).[26] For example, Fig. 2d shows $R_{12xy}= h/4e^2$ at B = 15T (where $m_1$ = 2, $m_2$ = 4 with ACW) and $R_{12xy}=- h/2e^2$ at B = -15T (where $m_1$ = -2, $m_2$ = -4 with CW), as predicted by Eq. (3) and Fig. 4c. We also note that the observed interface QHE could depend on the electrode pair used in the measurements. Instead of using "a" and "b" as the Hall electrodes, the measured quantized Hall resistance between electrodes "c" and "e" will show reversed switching behavior (see Fig. 4c).

Our analysis can also be extended to device "2". From Fig. 4a,b, we can get the geometry for device "2" by moving the probe "$c$" leftward (probe $c'$) to cover the interface (i. e. touching both SLG and BLG parts, as in devices "2"). The voltage $V_C$ measured by probe "c" is unchanged by this movement. The configuration of device "2" (electrodes "$S, D, c, b$" in Fig. 3a, measuring $R_{12xy}$) can be realized by electrodes "$e, c', S, D$" in Fig. 4a,b. Using the Onsager relation that holds for 4-terminal measurements in the linear response regime:

$$R_{SD,ec'}(CW/ACW) = R_{ec',SD}(ACW/CW) = R_{12xy}(ACW/CW) \qquad (4)$$

From the Hall resistance, $R_{SD,ec'} = \frac{V_e - V_{c'}}{I_{SD}} = -R_{SD,ce}$ (Eq. (3b)), we can get:

$$R_{12xy}(ACW) = -R_{SD,ce}(CW) = \frac{h}{|m_2|e^2}$$

$$R_{12xy}(CW) = -R_{SD,ce}(ACW) = -\frac{h}{|m_1|e^2} \quad (5)$$

These results are also consistent with the better-developed interface QH plateau resistances $R_{12xy}$ of device "2" (see Fig. 3c,d).

The theoretical plot in Fig. 4c is drawn assuming the SLG and BLG parts have the same and uniform Landau filling factor (*v*). However, in real samples, spatially non-uniform doping and/or charge transfer between the SLG and BLG parts may result in spatially nonuniform charge carrier densities, thus different or nonuniform *v* in SLG and BLG regions. Nonetheless, similar model and calculations as presented above can still apply whenever the SLG and BLG QHE plateaus overlap (which, for example, is the case for the well-developed interface QHE states observed in device "1"). On the other hand, if the QH states in the two regions do not appear simultaneously, then the interface QH states will not be well-developed. In addition, complicated edge state configurations at the interface have been predicted, arising from the interface LL structures (which could depend sensitively on interface orientation and disorder) [15]. These factors may result in less well developed interface QH states and a "mixed" appearance of both SLG-like and BLG-like QH plateaus in $R_{12xy}$ (e.g. in Fig. 3d) observed in some of our devices (particularly at lower *B*, where LLs are less resolved).

We also note that hybrid QH devices have been previously realized in gate-defined p-n junctions [24, 27-31] in pure SLG or BLG, where local top gates are used to create regions with different carrier densities (and/or types), thus different quantum Hall states [24, 27-31] (with different filling factors) in a magnetic field. The QHE observed in such *p-n* junctions has been successfully explained by a Landauer-Büttiker analysis of the edge states and their transmission or equilibration at the gate-defined interface between two regions with different quantum Hall states.[24, 29] Asymmetric (with *B* or carrier types) longitudinal and Hall resistances in the quantum Hall regime have also been observed in a SLG p-n junction using a four-terminal configuration [24] instead of two-terminal configurations [27-29], consistent with the Onsager relation. However, in a gated p-n junction, the gate-defined interface (e.g. its position as well as sharpness), thus where and how the edge-states from different regions meet and equilibrate, generally *vary* with the gate voltage. This can result in deviations of the QH plateaus from ideal quantized values (and from expected dependence on gate-tuned carrier densities), as often observed in experiments [24,27,28,30]. In our SLG-BLG junctions, the two regions with different QH states are created "naturally" because their *LL structures are different* (even when the two regions have the *same* carrier density and filling factor), allowing us to realize a QH junction with "intrinsic" interface (defined by the edge of the 2$^{nd}$ layer graphene lattice in the BLG region) and without the need of local gates (in contrast to p-n junctions in SLG or BLG [24,27-32]). The "natural" interface between the SLG and BLG is fixed at the edge of the BLG lattice and more sharply defined (down to atomic scale), making the SLG-BLG structure a potentially cleaner playground to study interface QHE in hybrid junctions. Furthermore, interface and junction QHE (dependent on the transmission and equilibration properties of edge states) can be a powerful tool to study QH edge physics [27-29]. Further improving our sample quality (eg., using boron nitride as the substrate) may allow us to probe many interesting questions regarding the QH edge physics in the regimes of broken-symmetry QHE [33-35] or fractional QHE (FQHE) [36,37]. Additional and more complex device structures, such as those with extra local gates, multi-segment planar junctions (eg. SLG-BLG-SLG), or junctions involving multilayer graphene (eg. trilayer graphene [38,39]) may also be envisioned, offering rich opportunities to study interface QHE and QH edge physics in hybrid structures involving many different electronic and LL configurations.

In conclusion, we have studied the QHE of the graphene planar hybrid structures consisting of partially SLG and BLG. The interface Hall resistance exhibits quantized plateaus where the normal electron-hole and magnetic field symmetries are no longer held. Furthermore, the interface quantized Hall resistances switch between those characteristic of SLG QH stats and BLG states when either the type of charge carriers or the direction of magnetic field is revered. A Landauer-Büttiker analysis is used to explain the observed SLG-BLG interface QH states, which are dependent only on the chirality of the edge states. Our work offers a new system to study the physics of junction QHE in graphene hybrid structures.

We acknowledge partial financial support from NSF, DTRA, Miller Family Endowment and Midwest Institute for Nanoelectronics Discovery (MIND). Part of the work was carried out at the National High Magnetic Field Laboratory, which is supported by NSF and the State of Florida. We thank G. Jones, T. Murphy, J.-H. Park and E. Palm for experimental assistance. Y. J. acknowledges support from NSF of China (Grants No. 11004174).

**References:**

1. K. S. Novoselov, A. K. Geim, S. V. Morozov, D. Jiang, Y. Zhang, S. V. Dubonos, I. V. Grigorieva, and A. A. Firsov, Science **306**, 666 (2004).

2. Y. B. Zhang, Y. W. Tan, H. L. Stormer, and P. Kim, Nature 4**38**, 201 (2005).

3. K. S. Novoselov, A. K. Geim, S. V. Morozov, D. Jiang, M. I. Katsnelson, I. V. Grigorieva, S. V. Dubonos, and A. A. Firsov, Nature **438**, 197 (2005).

4. A. K. Geim, Science. **324***,* 1530 (2009).

5. A. H. Castro Neto, F. Guinea, N. M. R. Peres, K. S. Novoselov, and A. K. Geim, Rev. Mod. Phys. **81***,* 109 (2009).

6. A. S. Mayorov, R. V. Gorbachev, S. V. Morozov, L. Britnell, R. Jalil, L. A. Ponomarenko, P. Blake, K. S. Novoselov, K. Watanabe, T. Taniguchi, and A. K. Geim, Nano Lett. **11**, 2396 (2011).

7. X. Du, I. Skachko, F. Duerr, A. Luican, and E. Y. Andrei, Nature **462**, 192, (2009).

8. K. I. Bolotin, F. Ghahari, M. D. Shulman, H. L. Strormer, and P. Kim, Nature **462**, 196, (2009).

9. C. R. Dean, A. F. Young, I. Meric, C. Lee, L. Wang, S. Sorgenfrei, K. Watanabe, T. Taniguchi, P. Kim, K. L. Shepard, and J. Hone, Nature Nanotech. **5**, 722 (2010).

10. K. S. Novoselov, E. McCann, S. V. Morozov, V. I. Fal'ko, M. I. Katsnelson, U. Zeitler, D. Jiang, F. Schedin, and A. K. Geim, Nature Phys. **2**, 177 (2006).

11. J. Velasco Jr, L. Jing, W. Bao, Y. Lee, P. Kratz, V. Aji, M. Bockrath, C. N. Lau, C. Varma, R. Stillwell, D. Smirnov, F. Zhang, J. Jung, and A. H. MacDonald, Nature Nanotech. **7**. 156 (2012).

12. P. Maher, C. R. Dean, A. F. Young, T. Taniguchi, K. Watanabe, K. L. Shepard, J. Hone, and P. Kim Nature Phys. **9**, 154 (2013).

13. R. E. Prange, S. M. Girvin, eds., *The Quantum Hall Effect*, 2$^{nd}$ edition. (Springer-Verlag, New-york, 1990).

14. C. P. Puls, N. E. Staley, and Y. Liu, Phys. Rev. B **79**, 235415 (2009)

15. M. Koshino, T. Nakanishi, and T. Ando, Phys. Rev. B **82**, 205436 (2010)

16. X. Xu, N. M. Gabor, J. S. Alden, A. M. van der Zande, and P. L. McEuen, Nano Lett. **10**, 562 (2010).

17. A. Tsukuda, H. Okunaga, D. Nakahara, K. Uchida, T. Konoike, and T. Osada, J. Phys.: conf. Ser. **334**, 012038 (2011).

18. Y. Zhao, PhD thesis, Columbia University, 2012.

19. J. F. Tian, L. A. Jauregui, G. Lopez, H. Cao, and Y. P. Chen, Appl. Phys. Lett. **96**, 263110 (2010).

20. P. Blake, E. W. Hill, A. H. Castro Neto, K. S. Novoselov, D. Jiang, R. Yang, T. J. Booth, and A. K. Geim, Appl. Phys. Lett. **91**, 063124 (2007).

21. A. C. Ferrari, J. C. Meyer, V. Scardaci, C. Casiraghi, M. Lazzeri, F. Mauri, S. Piscanec, D. Jiang, K. S. Novoselov, S. Roth, and A. K. Geim, Phys. Rev. Lett. **97**, 187401 (2006).

22. S. Datta, *Electronic transport in mesoscopic systems* (Cambridge University Press 1995)

23. T. Ihn, *Semiconductor Nanostructures* (Oxford, 2009)

24. D.-K. Ki, S.-G. Nam, H.-J. Lee, and B. Özyilmaz, Phys. Rev. B **81**, 033301 (2010)

25. In Fig. 3a,b and Eqs. (2-3), we assume that $|m_1|<|m_2|$ ($m_1 \cdot m_2 > 0$). Our

theoretical analysis can also be adapted to $|m_1|>|m_2|$ ($m_1 \cdot m_2 > 0$) by rotating the device in Figs. 4a,b by 180°. The calculated Hall resistances are then $R_{SD,ab} = h/(\max(m_1,m_2)e^2)$, $R_{SD,ce} = h/(\min(m_1,m_2)e^2)$ and longitudinal resistances are

$$R_{SD,ca} = \begin{cases} 0, CW \\ (1/m_2 - 1/m_1)h/e^2, ACW \end{cases}, \quad R_{SD,be} = \begin{cases} (1/m_1 - 1/m_2)h/e^2, CW \\ 0, ACW \end{cases}.$$

26. We point out that for Fig. 2c, the $-h/2e^2$ plateau observed in $R_{12xy}$ (near $V_g$=22V, where the SLG part has a QHE with $m_1$ = -2 and CW edge chirality, but the BLG part is near its CNP and not in a QH state) is not rigorously described by our simple model in Fig. 4ab. However, this plateau may still be explained by a Laudauer-Büttiker model by considering the entire BLG near its CNP (showing a dissipative transport with finite $R_{2xx}$) to act as part of the source ("S") contact for the SLG, resulting in $R_{12xy} = R_{1xy} = -h/2e^2$.

27. J. R. Williams, L. Dicarlo, and C. M. Marcus, Science, **317**, 638 (2007).

28. B. Özyilmaz, P. Jarillo-Herrero, D. Efetov, D. A. Abanin, L. S. Levitov, and P. Kim, Phys. Rev. Lett. **99**, 166804 (2007).

29. D. A. Abanin, and L. S. Levitov, Science, **317**, 641 (2007).

30. D. K. Ki, and H. J. Lee, Phys. Rev. B **79**, 195327 (2009).

31. L. Jing, J. Velasco Jr., P. Kratz, G. Liu, W. Bao, Marc Bockrath, and C. N. Lau, Nano Lett., **10**, 4000 (2010).

32. G. Liu, J. Velasco, Jr., W. Bao, and C. N. Lau, Appl. Phys. Lett. **92**, 203103 (2008).

33. Y. Zhang, Z. Jiang, J. P. Small, M. S. Purewal, Y.-W. Tan, M. Fazlollahi, J. D. Chudow, J. A. Jaszczak, H. L. Stormer, and P. Kim, Phys. Rev. Lett. **96**, 136806 (2006).

34. Y. Zhao, P. Cadden-Zimansky, Z. Jiang, and P. Kim, Phys. Rev. Lett. **104**, 066801 (2010).

35. D. A. Abanin, P. A. Lee, and L. S. Levitov, Phys. Rev. Lett. **96**, 176803 (2006).

36. C. R. Dean, A. F. Young, P. Cadden-Zimansky, L. Wang, H. Ren, K. Watanabe, T. Taniguchi, P. Kim, J. Hone, and K. L. Shepard, Nature Phys. **7**, 693 (2011).

37. Z.-X. Hu, R. N. Bhatt, X. Wan, and K. Yang, Phys. Rev. Lett. **107**, 236806 (2011).

38. L. Zhang, Y. zhang, J. Camacho, M. Khodas, and I. Zaliznyak, Nature Phys. **7**, 953, (2011).

39. T. Taychatanapat, K. Watanabe, T. Taniguchi, and P. Harillo-Herrero, Nature Phys. **7**, 621, (2011).

**Figure captions**

FIG. 1 (color online) (a) Optical image of a signal layer graphene (SLG) and bilayer graphene (BLG) hybrid structure (exfoliated from graphite). (b) Representative Raman spectra of the SLG and BLG parts in device "1". The wavelength of the Raman excitation laser is 532 nm. The power of the laser is ~200 μW incident on the sample. (c,d) The field effect (FE) curves (R vs. $V_g$) measured from the SLG ($R_{1xx}$) and BLG ($R_{2xx}$) parts for (c) device "1" and (d) device "2".

FIG. 2 (color online) (a) Optical image of devices "1". The contours for the regions corresponding to SLG and BLG are highlighted by white dotted and blue dashed lines, respectively. The widths of all the electrodes are 1 μm. (b) Schematic 3D structure of the device, indicating electrical connections for various resistance measurements. The positive *B* direction (black arrow) points upward. (c) $R_{1xy}$, $R_{2xy}$, and $R_{12xy}$ as well as $R_{1xx}$, $R_{2xx}$, and $R_{12xx}$ as functions of $V_g$ at B = 15 T and T = 0.5 K. (d) Hall resistances ($R_{1xy}$, $R_{2xy}$, and $R_{12xy}$) and longitudinal resistances ($R_{1xx}$, $R_{2xx}$, and $R_{12xx}$) as functions of the *B* at 0.5 K and $V_g$ = 0 V. In (c) and (d), QH plateaus associated with *SLG* or *BLG* QH states are labeled by arrows and their quantum numbers (m, related to the plateau values $R_{xy}=h/(me^2)$) in green or purple color, respectively. The edge state chirality, CW or ACW, has been labeled near the representative QH plateaus observed in $R_{12xy}$ in both (c) and (d).

FIG. 3 (color online) (a) Optical image of devices "2". The widths of all the electrodes are 1 μm. (b) Schematic 3D structure of the device with the corresponding electrical connections. (c) $R_{12xy}$ (between electrodes "c" and "b") as a function of $V_g$ at B = ±18 T (black or blue curves, respectively). (d) $R_{12xy}$ as a function of the *B* at $V_g$ = 0 V (blue curve) and 30 V (black curve). In (c) and (d), QH plateaus corresponding to values associated with SLG or BLG QH states are labeled by arrows and quantum numbers in green or purple color, respectively.

FIG. 4 (color online) Schematic of the edge state currents with (a) CW and (b) ACW circulation in both region "1" (white) and region "2" (purple). (c) Schematic illustration of the QHE in SLG, BLG, and SLG-BLG interface. The QHE $R_{xy}$ of either SLG (solid line) or BLG (dashed line) is related to the corresponding number (m) of edge states in SLG or BLG by $m=(h/e^2)/R_{xy}$, where $h/e^2$ the resistance quantum. The calculated interface QHE $R_{xy}$ takes the SLG or BLG values depending on the pair of electrodes used and the edge state chirality. While it is well known that $1/R_{xy}$ shows a jump of *$4e^2/h$* for SLG and *$8e^2/h$* for BLG at ν = 0 (within a single-particle physics picture), it is interesting to note that the jump for the SLG-BLG interface is *$6e^2/h$*.

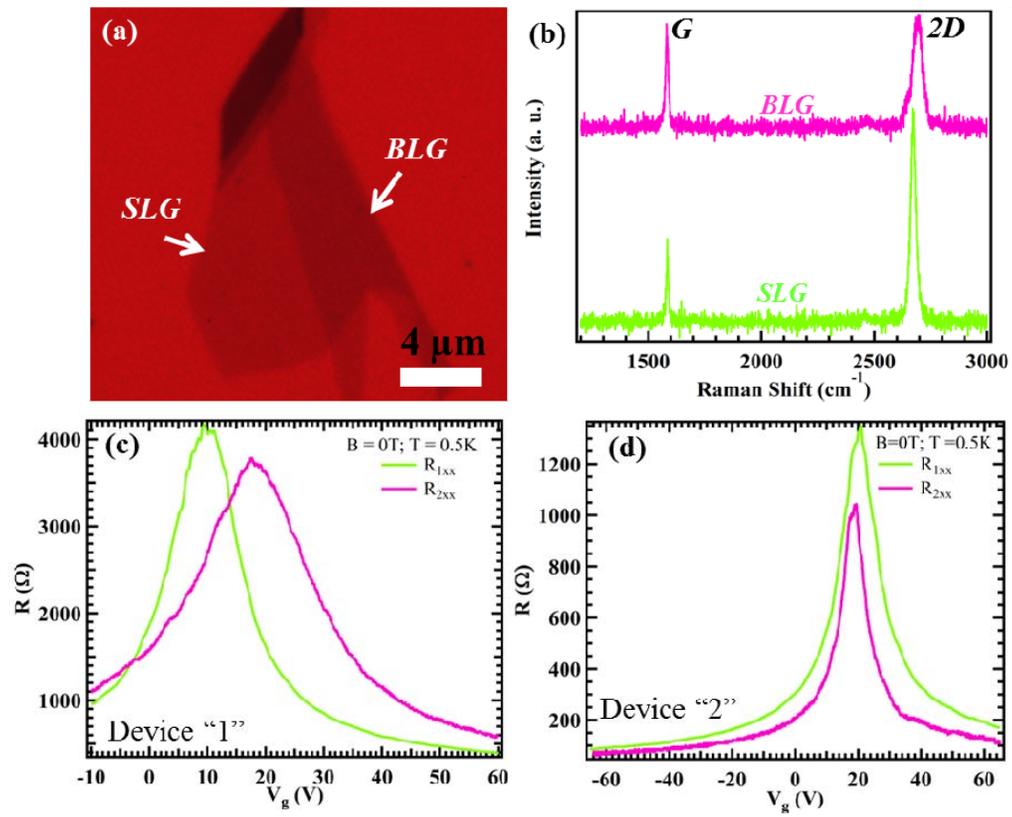

Figure 1 Tian *et al*.

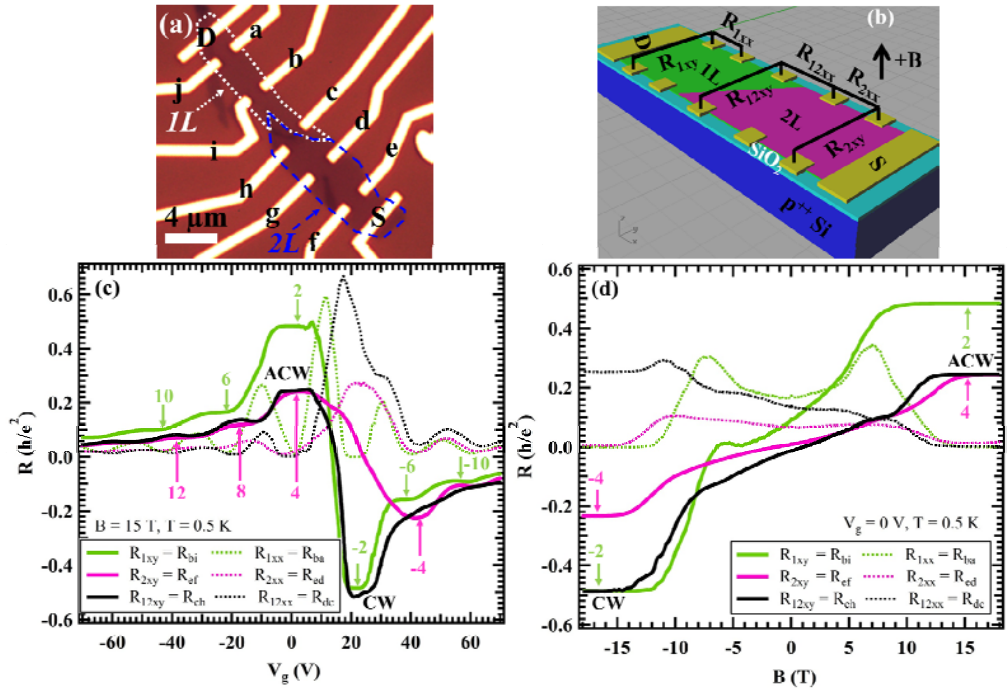

Figure 2 Tian *et al*.

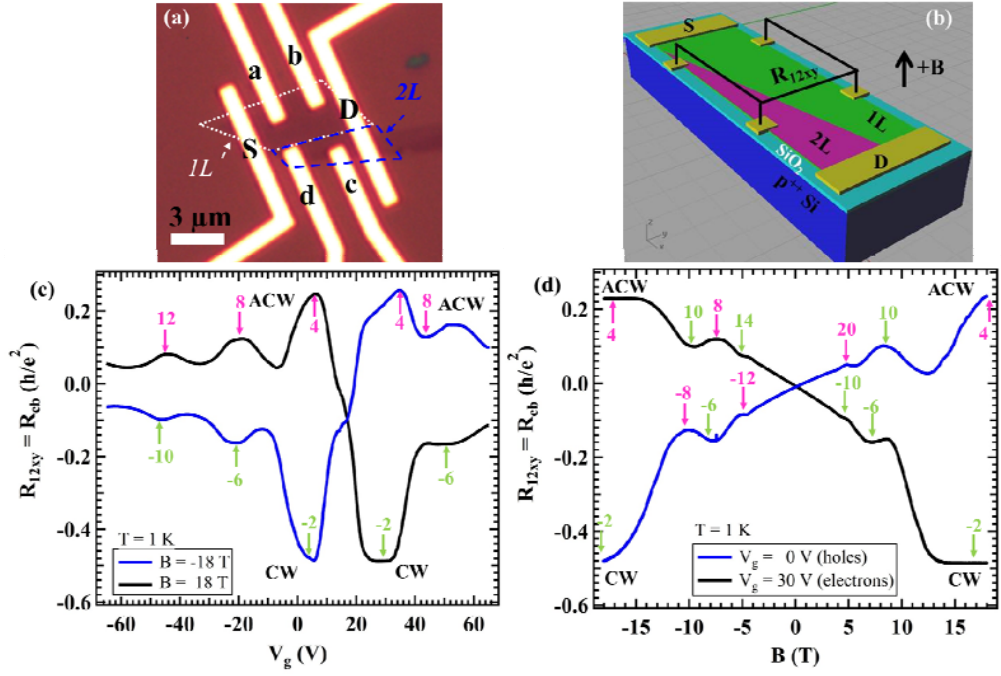

Figure 3 Tian *et al.*

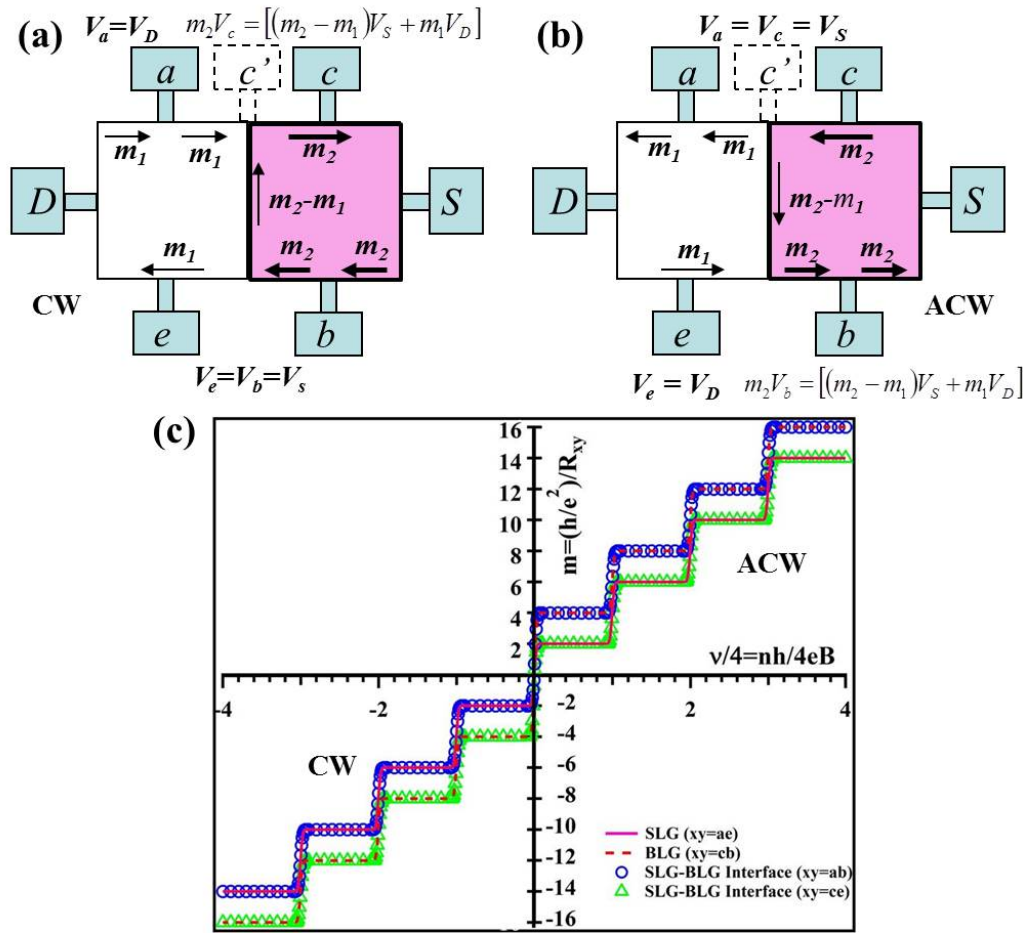

Figure 4 Tian *et al.*